\documentclass[conference]{IEEEtran}
\IEEEoverridecommandlockouts
\usepackage{cite}
\usepackage{comment}
\usepackage{amsmath,amssymb,amsfonts}
\usepackage{algorithmic}
\usepackage{graphicx}
\usepackage{textcomp}
\usepackage{xcolor}
\usepackage{multirow}
\usepackage{float}
\usepackage{tikz}
\usepackage{circuitikz}
\usepackage{caption}
\usepackage{svg}
\usepackage{subfig}
\usepackage{hyperref}
\usepackage{balance}

\def\BibTeX{{\rm B\kern-.05em{\sc i\kern-.025em b}\kern-.08em
    T\kern-.1667em\lower.7ex\hbox{E}\kern-.125emX}}

\hypersetup{
    colorlinks=true,
    linkcolor=black,
    urlcolor=black,
    filecolor=black,
    citecolor=black
}

\begin{document}

\author{\IEEEauthorblockN{Mohammed Elbtity, Peyton Chandarana, and Ramtin Zand}
\IEEEauthorblockA{Department of Computer Science and Engineering, University of South Carolina, Columbia, SC 29208, USA\\
e-mail: elbtity@ieee.org, psc@email.sc.edu, ramtin@cse.sc.edu
}
}

\bstctlcite{IEEEexample:BSTcontrol}

\title{Flex-TPU: A Flexible TPU with Runtime Reconfigurable Dataflow Architecture}

\maketitle
\begin{abstract}
Tensor processing units (TPUs) are one of the most well-known machine learning (ML) accelerators utilized at large scale in data centers as well as in tiny ML applications. TPUs offer several improvements and advantages over conventional ML accelerators, like graphical processing units (GPUs), being designed specifically to perform the multiply-accumulate (MAC) operations required in the matrix-matrix and matrix-vector multiplies extensively present throughout the execution of deep neural networks (DNNs). Such improvements include maximizing data reuse and minimizing data transfer by leveraging the temporal dataflow paradigms provided by the systolic array architecture. While this design provides a significant performance benefit, the current implementations are restricted to a single dataflow consisting of either input, output, or weight stationary architectures. This can limit the achievable performance of DNN inference and reduce the utilization of compute units. Therefore, the work herein consists of developing a reconfigurable dataflow TPU, called the Flex-TPU, which can dynamically change the dataflow per layer during run-time. Our experiments thoroughly test the viability of the Flex-TPU comparing it to conventional TPU designs across multiple well-known ML workloads. 
The results show that our Flex-TPU design achieves a significant performance increase of up to $2.75\times$ compared to conventional TPU, with only minor area and power overheads.

\end{abstract}

\section{Introduction}
\label{sec:introduction}


In 2015, Google launched its tensor processing unit (TPU) project 
adopting the systolic array architecture, dating back to as early as 1979 \cite{kung1979systolic}, to accelerate machine learning (ML) workloads \cite{jouppi2017quantifying,jouppi2017datacenter}. 
The first version of Google's TPU was primarily designed to accelerate ML workloads in large data centers
utilizing 8-bit integer (INT8) multiply and accumulate (MAC) units to offer a peak performance of 92 tera operations per second (TOPS) \cite{jouppi2017datacenter}. 
The most recent version of the  TPU, the TPU v4, can accelerate training and inference using the TPU's internal 16-bit brain-float (BF16) and INT8 precisions to offer up to 275 teraflops of computational power \cite{jouppi2023tpu}. In 2019, Google launched a smaller and low-power version of the TPU, called the Coral Edge TPU, that is suited to accelerate the inference of the ML workloads at the edge \cite{reidy2023efficient,reidy2023realtime,FERNeuro,edge_FER}. The Edge TPU uses INT8 MAC core units \cite{tf_models_on_edgetpu} and realizes a peak performance of four TOPS. 

In contrast to the graphical processing units (GPUs) already employed for this task, TPUs were specifically designed to accelerate the common matrix-matrix and matrix-vector multiplications dominant in ML workloads with a particular focus on maximizing data reuse while minimizing data transfer.
Since their advent in 2015, several variations of TPU have been proposed \cite{COSA,APTPU,pandey2021uptpu, hsu2021accelerating, TPU-IMAC,jouppi2023tpu}, which similarly adopt the systolic array architecture but focus on modifying the microarchitecture to achieve improvements in performance, power, energy, etc. 
For example, in 2022 the APTPU \cite{APTPU} was proposed which leverages approximate multipliers and adders in the systolic array processing elements (PEs) to improve the performance, area, and power of the TPU design. Another example is UPTPU \cite{pandey2021uptpu}, which utilizes power-gating to reduce the energy consumption of the TPUs.

\begin{figure}[]
\centering
\subfloat[]{
    \includegraphics[width=0.95\columnwidth]{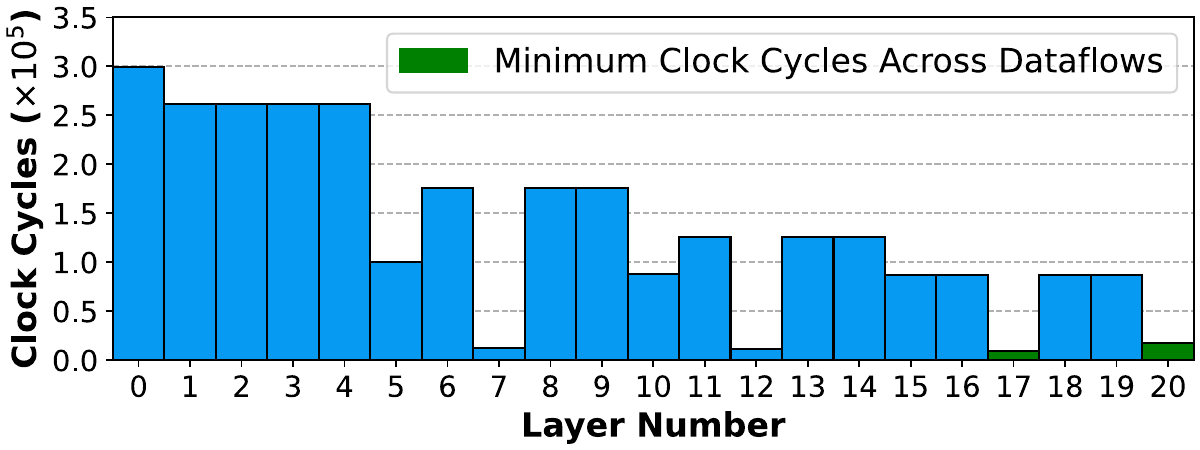}
}\\ 
\subfloat[]{
    \includegraphics[width=0.95\columnwidth]{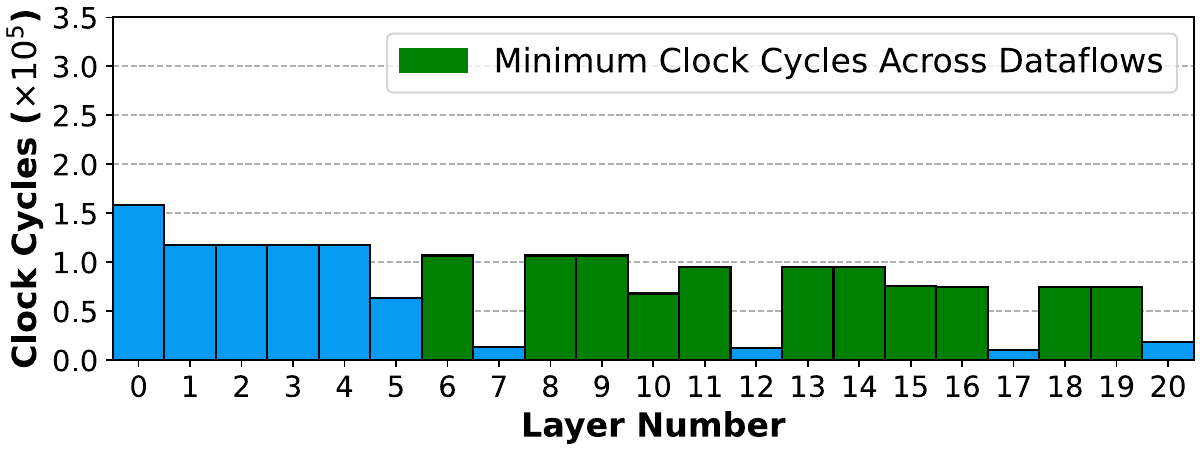}
}\\ 
\subfloat[]{
    \includegraphics[width=0.95\columnwidth]{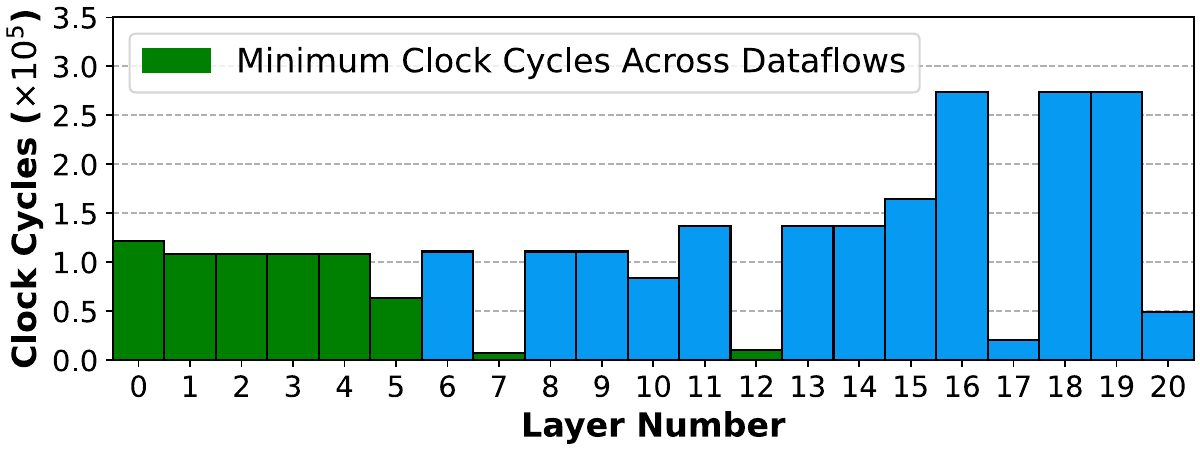}
}
\caption{Cycles required for executing each layer in ResNet-18 model using static dataflow architectures: (a) input stationary, (b) output Stationary, and (c) weight stationary. The layer-wise comparison shows that the optimal dataflow can be different in each layer of the network emphasizing the potential benefits of a flexible TPU with a run-time reconfigurable dataflow.
}
\label{fig:layers-dataflow}
\end{figure}

The typical systolic array architecture consists of an $N \times N$ array of PEs, each of which implements a MAC operation using a single multiplier and adder along with some registers to store data for reuse.
The dataflow in the systolic array is a mapping scheme that depends on the microarchitecture of PEs and determines how the input data is fed to the array along with how the partial results and outputs are generated and stored. 
Instead of loading and storing to and from memory for each computation, each PE in the systolic array typically employs one of the following dataflow paradigms: 

\begin{itemize}
    \item Input Stationary (IS): The inputs (or activations) remain fixed in the systolic array PEs while the weights are distributed horizontally. 
    
    \item Output Stationary (OS): Outputs are attached to MAC units as the inputs and weights are circulated among the units. As new inputs and weights are loaded and multiplied, they are then accumulated directly into the stationary outputs. 
    
    \item Weight Stationary (WS): Each weight is pre-loaded into a register attached to the MAC within each PE. During each cycle, the input activation data is multiplied by the fixed weights and broadcast across the systolic array's other processing elements.
\end{itemize}

Even in 2024, most of the TPUs used in larger data centers, or even the edge TPUs like the Google Coral Edge TPU \cite{edgeTPU}, have only been engineered with one of these three dataflow architectures in hardware. 
However, this singular static dataflow architecture may not always provide the optimal performance depending on the specific implementation of a deep neural network's (DNN) layers leading to significant performance limitations.
As shown in Fig. \ref{fig:layers-dataflow}, our simulations of the ResNet-18 \cite{resnet18} convolutional neural network (CNN), show that in many cases the many layers of a DNN can perform better on a heterogenous distribution of dataflows. For example in Fig. \ref{fig:layers-dataflow}, we see that ResNet-18's first five layers are fastest on the weight stationary dataflow while the more intermediate and final layers perform optimally on the output and input stationary dataflows, respectively.
As later shown in Fig. \ref{fig:layout}, a majority of the TPU's area is consumed by the systolic array and accounts for 77\%-80\% of the entire TPU's area consuming approximately 50\%-89\% of the overall power consumption of the architecture. Thus, modifying the microarchitecture of the systolic array to support multiple different dataflows could lead to significant performance speedups for the TPU as a whole.

In this paper, to increase the performance of the TPU, we propose a flexible runtime-reconfigurable dataflow TPU, called the Flex-TPU, in which the dataflow architecture of the systolic array can be reconfigured for each layer of the DNN according to the workload characteristic. Herein, our contributions consist of the following:
\begin{itemize}
    \item A modified PE microarchitecture to support runtime-reconfigurable dataflows.
    
    \item The implementation of the modified processing elements into a functional TPU.
    \item Thorough experimentation which shows the validity and increased performance of our design.
    
\end{itemize}

The remainder of the paper is organized as follows. In Section \ref{sec:proposed-arch} we present our Flex-TPU architecture and discuss 
the specific changes we make to the PE microarchitecture. 
In Section \ref{sec:results} we discuss the performance gains resulting from the flexibility in dataflow in the Flex-TPU and the marginal overheads incurred over conventional TPUs. Finally, we conclude the paper in Section IV.



\section{Proposed Flex-TPU}


The systolic array is the core element of any TPU architecture. A systolic array consists of two or multi-dimensional arrays of processing elements (PEs). Each PE in the systolic array implements a multiply-and-accumulate (MAC) operation by multiplying the weights and inputs with a multiplier and then adding this product with any previously computed partial sums using an accumulator. The result of this summation is then kept in the same PE or broadcast downstream to other PEs to be used in further computations. Regardless of the dataflow type, this MAC operation occurs in each PE of the systolic array to accomplish matrix-matrix or matrix-vector multiplication while maximizing data reuse without introducing additional data transfer overhead.

The primary distinction between different TPU architectures is typically the dataflow of the systolic array and its PEs. 
Each dataflow has its own advantages and trade-offs for ML workloads regarding power, data transfer, and compute units' utilization efficiency dependent on specific workload characteristics. The choice between the IS, OS, or WS dataflow largely depends on the objectives of the computation, such as maximizing data reuse, minimizing memory bandwidth, or reducing latency. As shown in Fig. \ref{fig:layers-dataflow}, ResNet-18's optimal dataflow varies across the layers between all three dataflows. Hence, selecting the optimal dataflow at runtime for each layer can lead to significant performance gains. 

\begin{figure}
    \centering
    \resizebox{0.95\columnwidth}{!}{
    \includegraphics{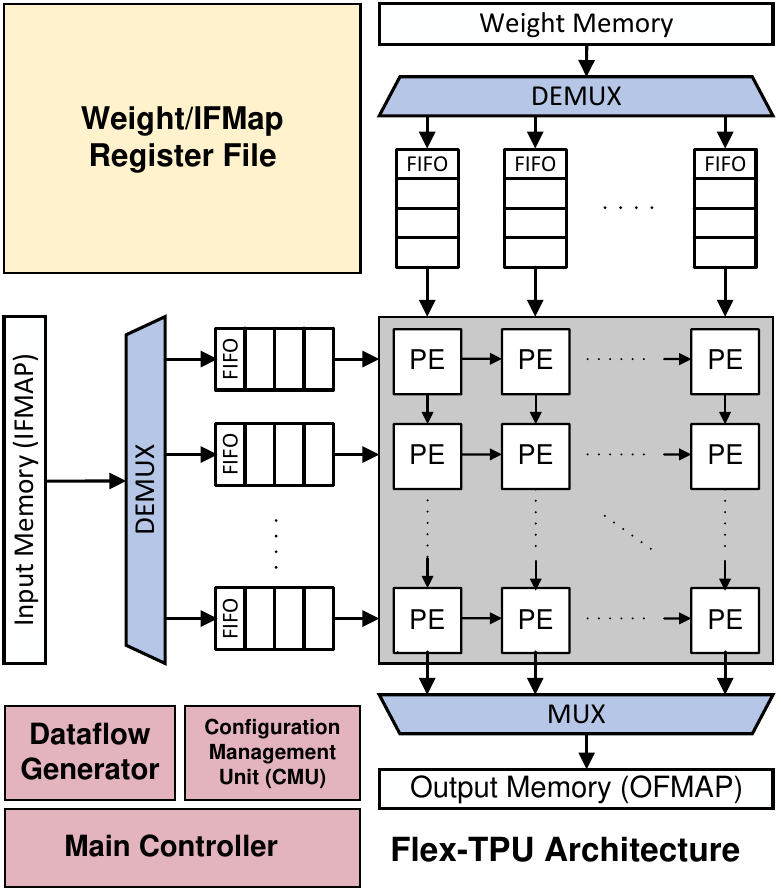}
    }
    \caption{The proposed Flex-TPU Architecture. 
    }
    \label{fig:tpu}
\end{figure}

Figure \ref{fig:tpu} shows the overall architecture of our proposed Flex-TPU, which is equipped with a runtime dataflow reconfigurability feature. Similar to a conventional TPU, our proposed Flex-TPU design consists of weight memory, input memory, output memory, and a systolic array of size $S=N \times N$ PEs surrounded by the first-in-first-out (FIFO) buffers as depicted in Fig. \ref{fig:tpu}. Additionally, our Flex-TPU includes a \textit{Weight/IFMap Register File} that stores the fixed or the ``stationary" weights or the IFMaps - depending on the selected dataflow - with output ports distributed among the PEs in the systolic array. Moreover, the \textit{Dataflow Generator} block generates the memory read/write addresses to store or retrieve the IFMaps, weights, and OFMap according to the selected dataflow dictated by the \textit{Configuration Management Unit (CMU)}. The CMU selects the dataflow for each layer of the ML workload by informing the \textit{Dataflow Generator} and by reconfiguring the PEs within the systolic array to work according to the pre-determined dataflow for each layer. 
It is worth mentioning that the dataflow of each layer of the ML model is determined after training and before deployment of the model on the Flex-TPU, reducing the complexity of the hardware. The \textit{Main Controller} handles the data transfer between memories/FIFOs and the systolic array, programming the CMU units, and writes to the \textit{Weight/IFMap Register File}. The proposed Flex-TPU's architecture differs from that of the conventional TPU in two primary ways: 1) the processing elements within the systolic array and 2) the controller driving the dataflow selections.

\label{sec:proposed-arch}

\begin{figure}
    \centering
    \resizebox{\columnwidth}{!}{
    \includegraphics{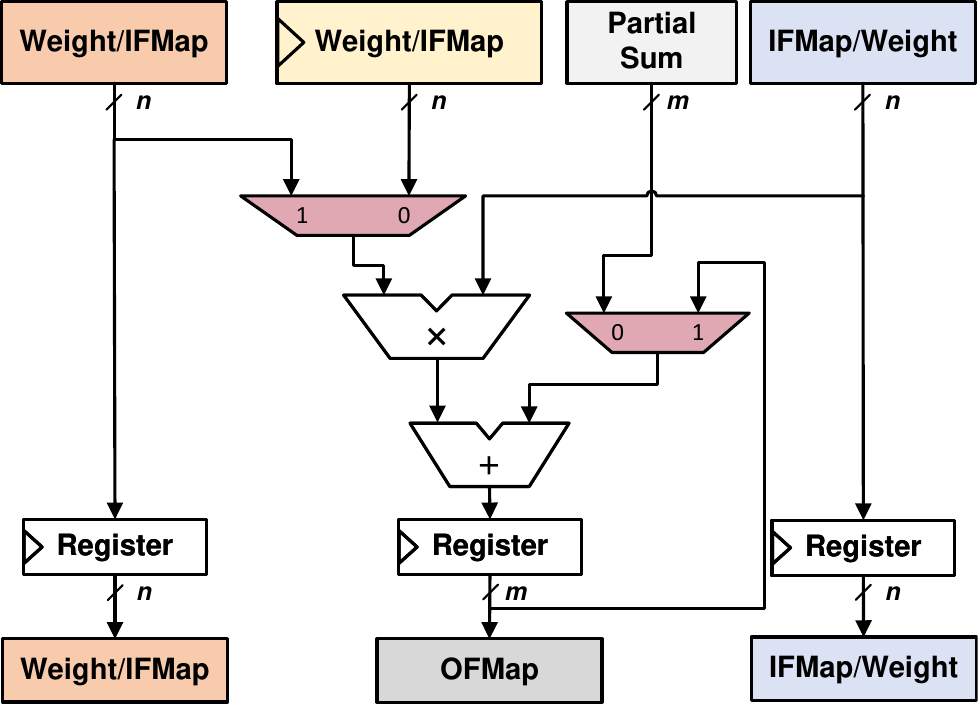}
    }
   \caption{The proposed Flex-TPU processing element (PE) with runtime reconfigurable dataflow.}
    \label{fig:flex-pe}
\end{figure}

\begin{figure*}
    \centering
    \subfloat[]{
    \includegraphics[width=.32\textwidth]{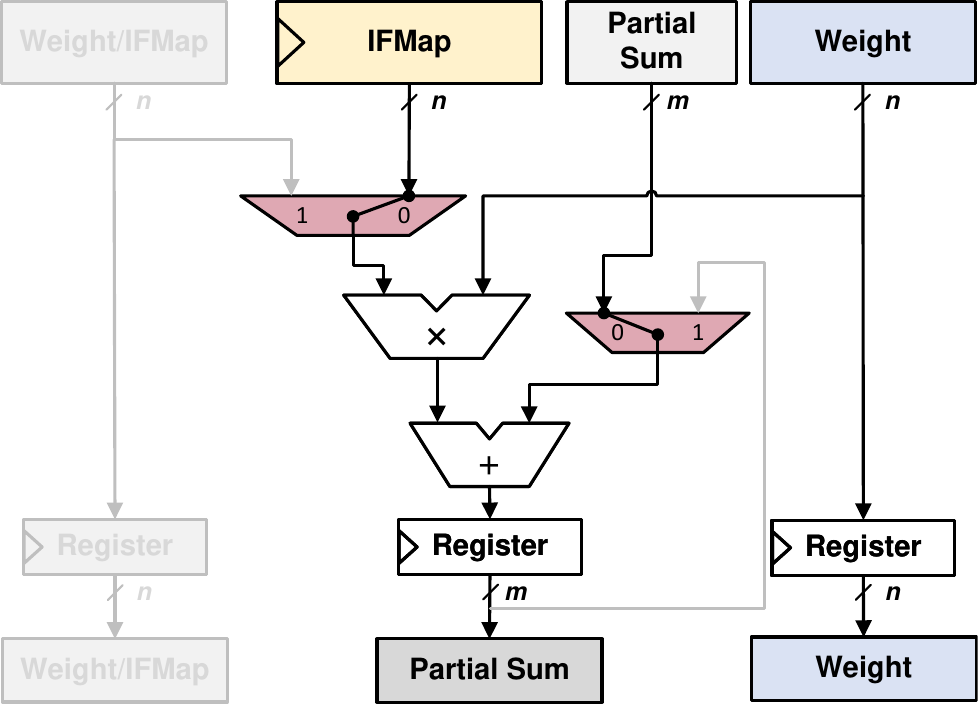}
    }
\subfloat[]{
    \includegraphics[width=.32\textwidth]{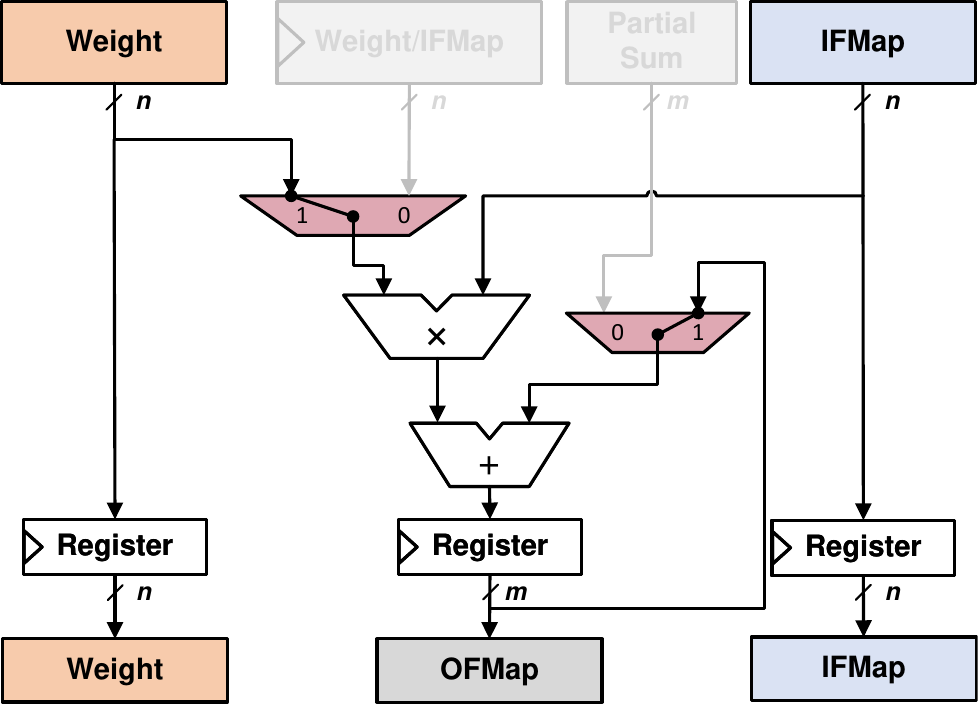}
    }
\subfloat[]{
    \includegraphics[width=.32\textwidth]{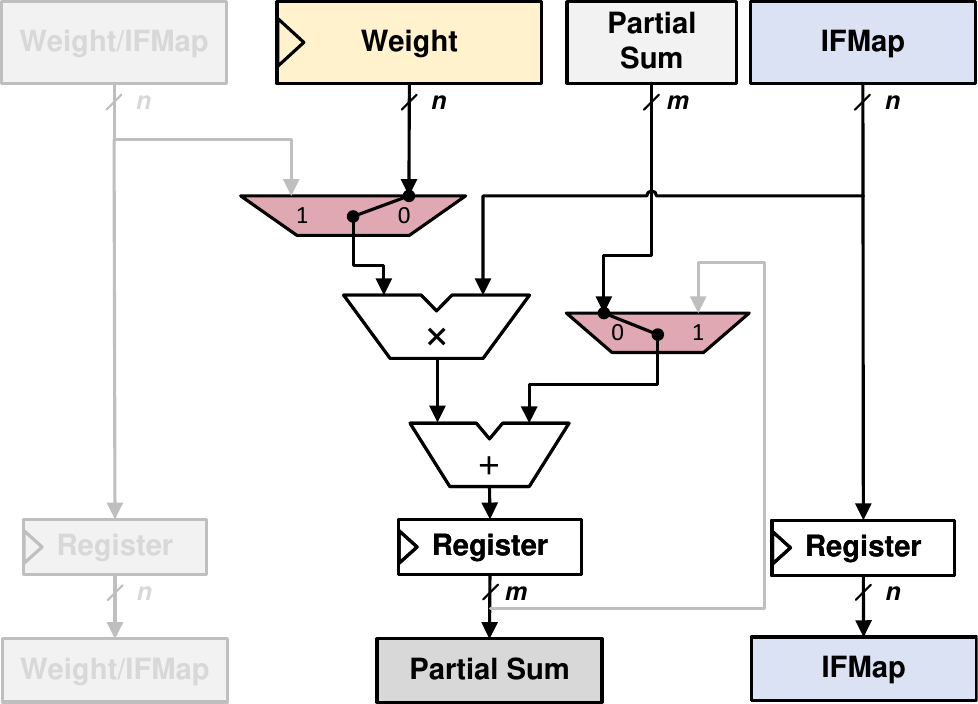}
    }
    \caption{The three flexible PE dataflow configurations controlled by the two added MUXs: (a) IS, (b) OS, and (c) WS modes.}
    \label{fig:flex-flow}
\end{figure*}

Figure \ref{fig:flex-pe} shows the microarchitecture of a processing element in the Flex-TPU, which has one extra register and two multiplexers (MUXs) compared to the PE of a conventional TPU. The MUXs are controlled by the configuration management unit and are utilized to select the optimal dataflow for each layer in the ML model during runtime. 
As investigated in Section \ref{sec:power-area-results}, adding these three extra components to each PE in the systolic array does moderately increase both area and power consumption, but the flexibility of the design provides a significant performance increase as discussed in section \ref{sec:performance-results}.

As shown in Fig. \ref{fig:flex-flow}, there are three possible runtime configurations of our flexible PEs that correspond to each of the three common dataflows: IS, OS, and WS. In Fig. \ref{fig:flex-flow} (a), the input stationary (IS) configuration is shown where the input-feature-map (IFMap) is fixed in a register in the PE. To accomplish the IS dataflow during runtime, the \textit{CMU} sends both MUXs a ``0'' control signal and the \textit{Main Controller} pins the IFMap in the register in the PE. 
The IS dataflow often excels for layers with small stridden convolutions or depthwise convolutions due to the high input reuse. By minimizing the movement of heavily reused input data, a significant amount of bandwidth can be saved, particularly in memory-bound operations or power-constrained environments.

Figure \ref{fig:flex-flow}(b) shows the OS configuration mode of the PE. In this mode, the IFMaps and Weights are multiplied and then moved through the PEs in the systolic array to be reused in further operations. The partial sums remain fixed in the PEs and keep accumulating to form the final output feature map (OFMap) of the layer. 
The OS dataflow mode is triggered by a ``1'' control signal being sent from the \textit{CMU} to the MUXs of each PE and thus leads to the output of the MAC being fixed inside of the accumulator. The OS dataflow is typically advantageous in deeper layers of DNNs where a large number of partial sums are being accumulated. Thus, keeping the output fixed within the PE minimizes the need for frequent memory stores of intermediate results benefiting layers with higher computational intensity per output.

Figure \ref{fig:flex-flow} (c) shows the runtime configuration for the weight stationary (WS) dataflow. 
The WS dataflow mode is activated with a ``0'' control signal sent from the \textit{CMU}. However, instead of the IFMap being fixed in the added register in the PE, the weight is fixed for the duration of the computation. This can be advantageous in the first layers of DNNs where the ratio of input to weights is high. As a result, keeping the weights stationary yields the efficient use of memory bandwidth and improves the overall computational throughput. 

Selecting the optimal dataflow strategy is dependent on multiple layer-specific characteristics such as IFMap dimensions, filter sizes, number of channels, and strides. 
To find the optimal dataflow strategy for each layer in the DNN, we should run each trained model on the Flex-TPU three times, once for each dataflow, during the development phase. 
From these three runs, the dataflow that executes each layer's computation in the least number of clock cycles is then selected as the optimal dataflow for that layer. Following this one-time pre-deployment optimization procedure the optimal dataflow is then programmed into the \textit{CMU} by the \textit{Main Controller} and the \textit{CMU} subsequently 
drives each processing element's MUXs to reconfigure them with 
the optimal dataflow during runtime as well as informing the \textit{Dataflow Generator} to generate the read/write indices accordingly. 
This process only needs to be performed once per DNN model prior to deployment and during the development phase to optimize the per-layer dataflows.
While not a part of this work, we plan to explore other methods of selecting the optimal dataflow for networks deployed on the Flex-TPU in the future.

\section{Results and Discussions}
\label{sec:results}

\begin{figure}[]
    \centering
    \resizebox{.6\columnwidth}{!}{
    \includegraphics{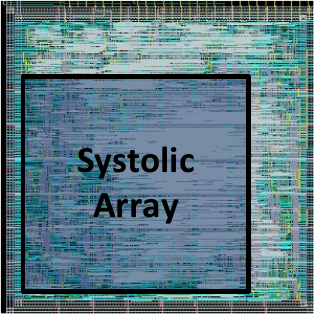}
    }
    \caption{The layout of the in-house designed TPU chip exhibiting the ratio of the systolic array compared to the surrounding logic and controller.}
    \label{fig:layout}
\end{figure}

As mentioned in Section \ref{sec:proposed-arch}, the proposed Flex-TPU design modifies the processing elements in the systolic array architecture to introduce dataflow flexibility with the inclusion of a single extra register and two multiplexers. While this work does not focus on making improvements to the other components of the architecture, like the FIFOs, the systolic array is estimated to utilize approximately 77\% to 80\% of the entire TPU's area, as shown in Fig. \ref{fig:layout}. Additionally, the systolic array also accounts for approximately 50\%-89\% of the power consumption depending on the systolic array size $S$. Thus, modifying the PEs of the systolic array can significantly affect the overall performance of the TPU. The layout of the in-house implementation of a TPU chip is shown in Fig. \ref{fig:layout} and demonstrates the overall ratio of the systolic array's area compared to the entire layout. 

Herein, our results in Section \ref{sec:performance-results} show the benefits of the Flex-TPU architecture in providing layer-by-layer dataflow optimizations to increase the overall throughput and performance of a systolic array of size $S=32\times32$. 
As mentioned before, while there are significant performance gains to the Flex-TPU design, this increased performance does contribute to a slight increase in area and power consumption that is discussed in Section \ref{sec:power-area-results}. The performance, power, and area results are obtained using ScaleSim V2 \cite{samajdar2018scale, samajdar2020systematic}, a cycle-accurate simulator for ML accelerators, and Synopsys Design Compiler along with the 45nm Nangate Open Cell Library.

\subsection{Performance}
\label{sec:performance-results}


\begin{table}[]
\centering
\caption{Clock cycles required for Flex-TPU versus conventional TPU with static dataflow.}
\label{tab:scalesim-table}
\resizebox{\columnwidth}{!}{%
\begin{tabular}{ccccc}
\hline
\textbf{Model} & \textbf{\begin{tabular}[c]{@{}c@{}}Flex-TPU\\ Cycles\end{tabular}} & \textbf{Dataflow} & \textbf{\begin{tabular}[c]{@{}c@{}}Static Dataflow\\ Cycles\end{tabular}} & \textbf{Speedup} \\ \hline
\multirow{3}{*}{\textbf{\begin{tabular}[c]{@{}c@{}}AlexNet\\ \cite{alexnet}\end{tabular}}} & \multirow{3}{*}{8.598e+5} & IS & 1.176e+6 & 1.368 \\ 
 &  & OS & 8.852e+5 & 1.030 \\ 
 &  & WS & 1.188e+6 & 1.382 \\ \hline
\multirow{3}{*}{\textbf{\begin{tabular}[c]{@{}c@{}}FasterRCNN\\ \cite{fasterrcnn}\end{tabular}}} & \multirow{3}{*}{3.922e+6} & IS & 5.640e+6 & 1.438 \\ 
 &  & OS & 4.368e+6 & 1.114 \\ 
 &  & WS & 4.710e+6 & 1.201 \\ \hline
\multirow{3}{*}{\textbf{\begin{tabular}[c]{@{}c@{}}GoogleNet\\ \cite{googlenet}\end{tabular}}} & \multirow{3}{*}{1.566e+6} & IS & 2.525e+6 & 1.612 \\ 
 &  & OS & 1.660e+6 & 1.060 \\ 
 &  & WS & 1.988e+6 & 1.269 \\ \hline
\multirow{3}{*}{\textbf{\begin{tabular}[c]{@{}c@{}}MobileNet\\ \cite{mobilenetv1}\end{tabular}}} & \multirow{3}{*}{1.206e+6} & IS & 2.349e+6 & 1.949 \\ 
 &  & OS & 1.373e+6 & 1.139 \\ 
 &  & WS & 1.531e+6 & 1.270 \\ \hline
\multirow{3}{*}{\textbf{\begin{tabular}[c]{@{}c@{}}ResNet-18\\ \cite{resnet18}\end{tabular}}} & \multirow{3}{*}{1.636e+6} & IS & 2.839e+6 & 1.736 \\ 
 &  & OS & 1.718e+6 & 1.051 \\ 
 &  & WS & 2.520e+6 & 1.540 \\ \hline
\multirow{3}{*}{\textbf{\begin{tabular}[c]{@{}c@{}}VGG-13\\ \cite{vgg19}\end{tabular}}} & \multirow{3}{*}{2.172e+7} & IS & 2.971e+7 & 1.368 \\ 
 &  & OS & 2.231e+7 & 1.027 \\ 
 &  & WS & 3.046e+7 & 1.402 \\ \hline
\multirow{3}{*}{\textbf{\begin{tabular}[c]{@{}c@{}}YOLO-Tiny\\ \cite{yolov4}\end{tabular}}} & \multirow{3}{*}{2.131e+6} & IS & 3.729e+6 & 1.750 \\ 
 &  & OS & 2.550e+6 & 1.196 \\ 
 &  & WS & 3.337e+6 & 1.566 \\ \hline
\end{tabular}%
}
\vspace{-4mm}
\end{table}

\begin{table*}[]
\centering
\caption{Area, power, and critical path delay overheads comparing the Flex-TPU to a conventional TPU with OS dataflow.}
\vspace{-1mm}
\begin{tabular}{cccc|ccc|ccc}
\hline
\multirow{2}{*}{S} & \multicolumn{3}{c|}{Area ($mm^2$)}  & \multicolumn{3}{c|}{Power ($mW$)}   & \multicolumn{3}{c}{Critical Path Delay ($ns$)}  \\ \cline{2-10} 
                   & TPU   & Flex-TPU & Overhead  & TPU    & Flex-TPU & Overhead  & TPU  & Flex-TPU & Overhead \\ \hline
$8\times8$               & 0.070 & 0.080    & 13.607\%        & 3.491  & 3.756    & 7.591\%         & 5.80 & 5.92     & 2.07\%          \\
$16\times16$             & 0.284 & 0.318    & 12.180\%        & 13.850 & 15.241   & 10.045\%        & 6.44 & 6.48     & 0.62\%          \\
$32\times32$             & 1.192 & 1.311    & 10.052\%        & 55.621 & 61.545   & 10.650\%        & 6.63 & 6.69     & 0.90\%          \\ \hline
\end{tabular}
\label{tab:area-power-delay}
\end{table*}

In our experiments, the optimal dataflow was found by running each of the selected DNN models on the three different dataflows, IS, OS, and WS, using the ScaleSim V2 systolic array simulator \cite{samajdar2018scale, samajdar2020systematic}. ScaleSim provides a layer-by-layer summary of the clock cycles required by a user-defined systolic array of size $S=N \times N$ to perform the inference computations of a specific DNN. Each of these DNNs is comprised of multiple convolutions and fully connected layers and is deployed on a TPU with $S=32 \times 32$ systolic array. 
Table \ref{tab:scalesim-table}, provides the total number of clock cycles required by each model using our Flex-TPU design as well as on a conventional TPU with a single static dataflow. As shown in Table \ref{tab:scalesim-table}, the Flex-TPU provides a speedup of 1.027$\times$ to 1.949$\times$ across all models and dataflows.

\begin{figure}[]
    \centering
    \resizebox{\columnwidth}{!}{
    \includegraphics{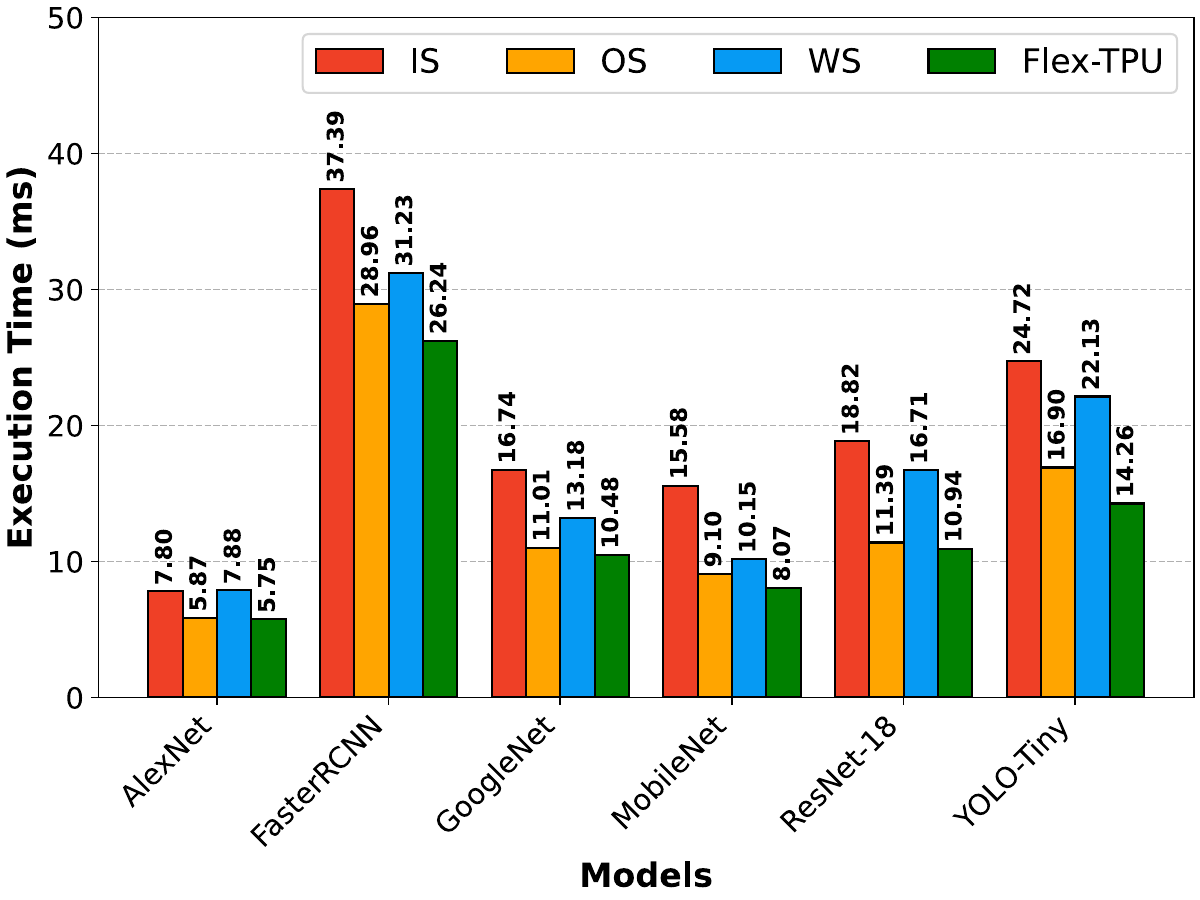}
    }
    \caption{The inference time per model for a systolic array size of $S=32 \times 32$ for the varying dataflows: IS, OS, WS, and our Flex-TPU. VGG is not shown because its notably longer execution time disrupts the clarity of the graph.}
    \label{fig:scalesim-barchart}
\vspace{-6mm}    
\end{figure}

Analyzing the speedup for each dataflow, most of the models perform close to optimally employing the OS dataflow compared to the IS and WS dataflows. Across the investigated models, the FasterRCNN and YOLO-Tiny appear to experience a greater speedup using the OS dataflow. On average, the Flex-TPU's reconfigurable dataflow provides a speedup of $1.612\times$, $1.090\times$, and $1.400\times$ compared to the single static dataflows of IS, OS, and WS, respectively. This implies that the IS and WS dataflows performed the poorest in terms of their overall computational execution time.

To determine the real-world execution time of each model running on a single dataflow, the number of clock cycles for each model was multiplied by the critical path delay of $6.63\ ns$ for the conventional $S=32 \times 32$ TPU. For the same $S=32 \times 32$ size configuration of the Flex-TPU, we multiplied the optimal number of clock cycles found per layer by the Flex-TPU's critical path delay of $6.69\ ns$ to obtain each model's execution time. The execution times across the tested modes for both the static dataflows and the Flex-TPU are shown in Fig. \ref{fig:scalesim-barchart}. Across all models, the Flex-TPU is the best architecture in terms of execution time, outperforming conventional TPU architecture with static dataflows by as much as $10.99\ ms$ which could be the determining factor of whether a model is considered real-time or not.

\subsection{Power and Area Overheads}
\label{sec:power-area-results}

To examine the area and power utilization of our Flex-TPU architecture, we employ the Synopsys Design Compiler to synthesize a conventional TPU design along with the Flex-TPU under the same design constraints. These constraints consist of an uncertainty of 2\%, a clock period of 10 $ns$, and a clock network delay of 1 $ns$. In particular, we synthesized three systolic array sizes consisting of $S = 8\times8$, $16\times16$, and $ 32\times32$ for both the conventional TPU and our proposed Flex-TPU and report the area, power, and critical path delay for each in Table \ref{tab:area-power-delay}. 
In this context, we focus solely on the OS dataflow architecture for the conventional TPU, as it achieves the best performance when compared to the IS and WS dataflow, as discussed in the previous subsection.

As expected, in Table \ref{tab:area-power-delay}, the area and power consumption of the Flex-TPU are marginally higher compared to that of the conventional TPU with an overhead of at most 13.607\% and 6.44\%, respectively. Considering the potential average speedup of 36.7\% achieved across all dataflows from Table \ref{tab:scalesim-table} and the improved execution times shown in Figure \ref{fig:scalesim-barchart}, this area and power overhead could be considered acceptable depending on the network to be deployed and considering applications where the utmost speed is desired in a system.

While both the area and power increase in the Flex-TPU, the critical path delay remains very similar across each of the systolic array sizes ($S$), not exceeding more than 2.07\% in the worst-case, i.e. $S=8\times8$. This implies that the Flex-TPU's architecture, if implemented in silicon, could have a very comparable clock frequency to the conventional TPU and further highlights the aforementioned speedups discussed as net speed increases in the Flex-TPU's overall performance.



\begin{figure*}[]
    \centering
    \subfloat[]{
    \includegraphics[width=.45\textwidth]{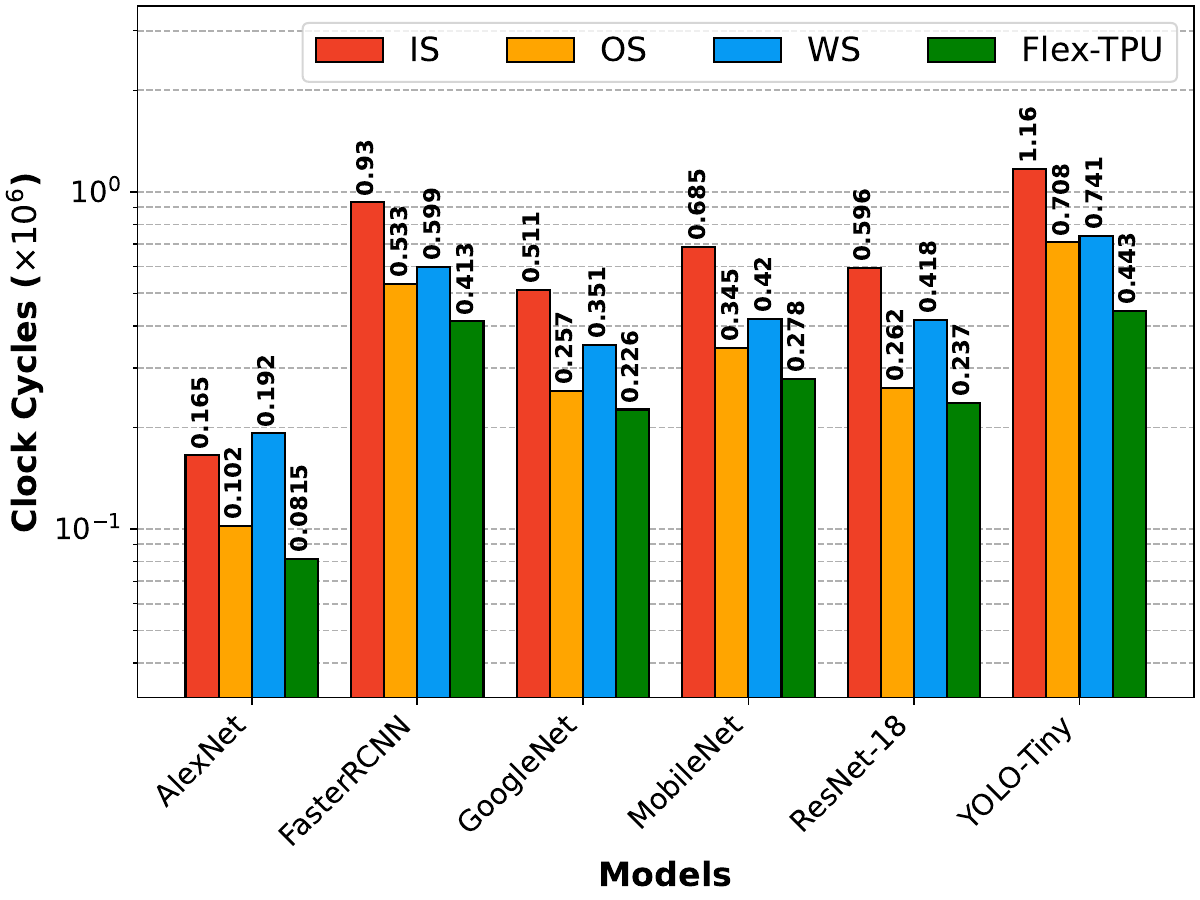}}
    \subfloat[]{
    \includegraphics[width=.45\textwidth]{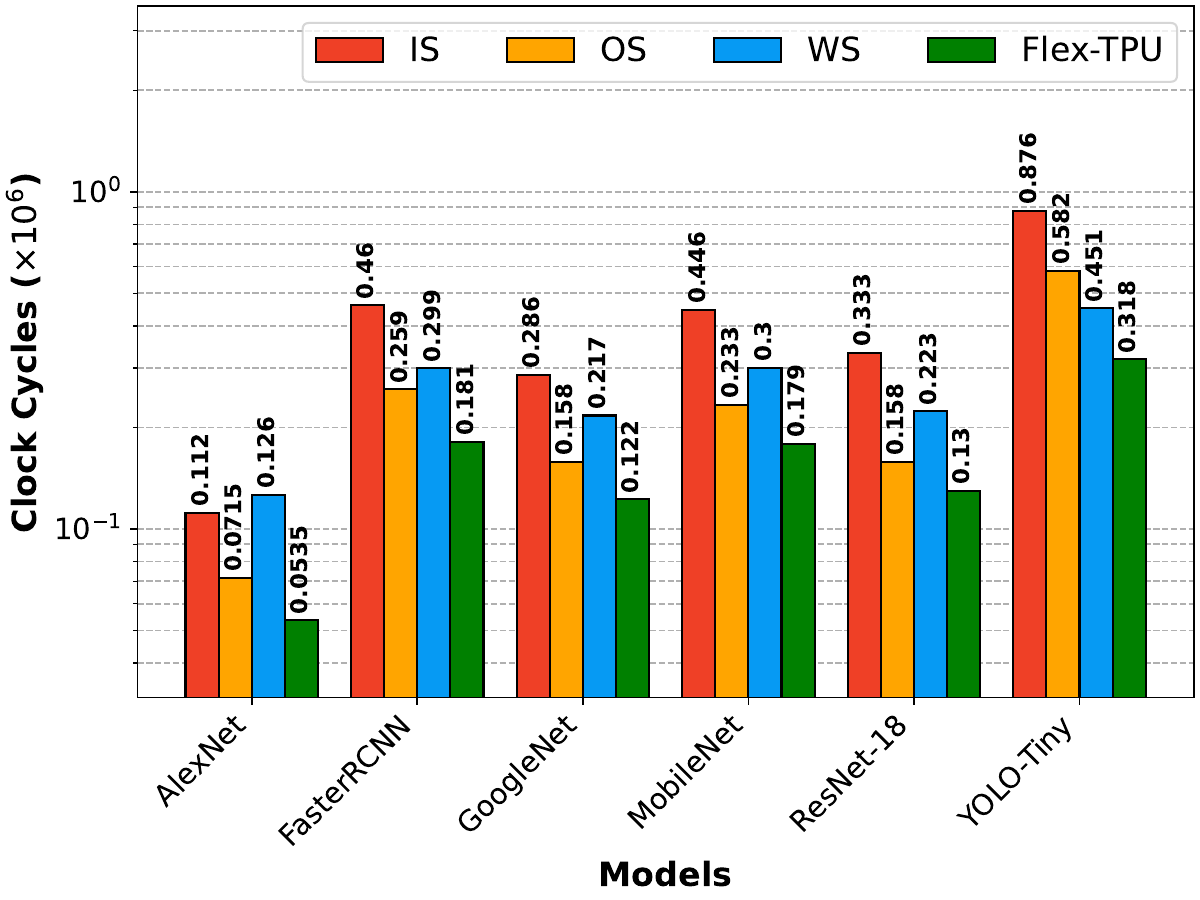}}
    \vspace{-1mm}
    \caption{The inference clock cycles per model for a systolic array sizes of (a) $S=128 \times 128$ and (b) $S=256 \times 256$ for the varying dataflows: IS, OS, WS, and our Flex-TPU demonstrating the scalability of our proposed Flex-TPU architecture.}
    \label{fig:scalesim-barchart-scaled}
\vspace{-3mm}    
\end{figure*}

\subsection{Scalability}

Thus far, our Flex-TPU architecture results have consisted of small systolic array sizes of $S=8 \times 8$, $16 \times 16$, and $32 \times 32$. These systolic array sizes would typically be used in smaller devices such as the Google Coral Edge TPU \cite{coral_inference_overview} since they are tailored for applications at the edge. 
In this subsection, we focus on investigating the current data center scale TPUs like the Google TPU v1 \cite{jouppi2017datacenter} with a $256\times256$ systolic array. 

In Figure \ref{fig:scalesim-barchart-scaled}, we compare the TPU designs with static dataflow against our Flex-TPU architecture at a larger systolic array size of $S=128 \times 128$ and $256 \times 256$. 
Similar to the smaller scale $S=32 \times 32$ systolic array, the Flex-TPU still provides a significant speed advantage compared to the IS and WS dataflow. However, compared to the TPU with OS dataflow, the Flex-TPU achieves further performance gains at scale.  
In particular, the Flex-TPU with the $128 \times 128$ systolic array achieves an average speedup of  1.238$\times$, and the $256 \times 256$  achieves a 1.349$\times$ speedup compared to the 1.090$\times$ speedup of the $32 \times 32$ systolic array. This demonstrates the Flex-TPU's effectiveness in further accelerating ML workloads in data centers at a larger scale.


\section{Conclusion}


As AI and ML gain increasing traction in daily life, there is a rising demand for more performant systems and architectures for accelerating these workloads. While conventional TPUs have been instrumental in keeping up with this demand, their current static dataflow implementation potentially inhibits their full potential on some workloads. Thus, selecting an optimal dataflow specific to the workload can lead to significant performance gains. Herein, we proposed the Flex-TPU architecture highlighting the potential to further optimize the TPU design's performance without limitations caused by static dataflows. The Flex-TPU accomplishes a runtime-reconfigurable dataflow by adding two multiplexers and a single register to each processing element. 
The experiments and simulation results demonstrate performance increases across various ML workloads with up to a 2.75$\times$ speedup without incurring significant area and power overheads. Considering the popularity of current TPU accelerators in data centers and edge applications, the higher performance achieved by the Flex-TPU positions it as an appealing upgrade for future implementations of TPUs.

\section*{Acknowledgment}
This work is supported by the National Science Foundation (NSF) under grant number 2340249.

\balance
\bibliographystyle{IEEEtran}
\bibliography{refs}

\end{document}